\documentclass[aps,prl,reprint,groupedaddress,amsmath]{revtex4-2}

\usepackage[colorlinks,linkcolor=blue]{hyperref}
\usepackage{graphicx}
\usepackage{dcolumn}
\usepackage{bm}
\usepackage{footnote}
\usepackage{extarrows}

\begin{document}


\title{Geometric Approach to Circular Photon Orbits and Black Hole Shadows}


\author{Chen-Kai Qiao}
\email{chenkaiqiao@cqut.edu.cn}
\affiliation{College of Science, Chongqing University of Technology, Banan, Chongqing, 400054, China}

\author{Ming Li}
\email{mingli@cqut.edu.cn}
\affiliation{Mathematical Science Research Center, Chongqing University of Technology, Banan, Chongqing, 400054, China}


\date{\today}

\begin{abstract}
Circular photon orbit and black hole shadow are significantly important issues in physics and astronomy, and a number of breakthroughs have been witnessed in recent years. Conventionally, the stable and unstable circular photon orbits are obtained using the effective potential of test particles moving in black hole spacetime. In this work, a pure geometric approach is developed to calculate these circular photon orbits and black hole shadow radius. Furthermore, it can be proved that our geometric approach is completely equivalent to the conventional approach based on effective potentials of test particles.




\end{abstract}


\maketitle

\textbf{Backgrounds and Introduction:}
Black holes are massive compact objects predicted by Einstein's general theory of relativity. They have attracted large numbers of interests in high-energy physics, astrophysics and astronomy over the past decades. Significantly important information on gravitation, galaxies, thermodynamics and quantum effects in curved spacetime can be revealed from black holes \cite{Hawking1974,Witten1998,Ferrarese2000,Ryu2006,Black Hole Physics}. Recently, huge progresses in black hole physics have been witnessed. The gravitational wave signals from binary black hole mergers were detected by LIGO and Virgo \cite{LIGO2016,LIGO2016b}. The high resolution images of supermassive black hole at the center of galaxy M87 were captured by Event Horizon Telescope (EHT) \cite{ETH2019a,ETH2019b}.

The circular photon orbit and shadow radius are important features for black holes. The particle motions, gravitational lensing, optical imaging and other aspects of black hole can be studied from these quantities. Since the observation of black hole image in galaxy M87 by EHT collaboration, the circular photon orbits and black hole shadow have became extremely hot topics in physics and astronomy. Conventionally, these circular photon orbits and black hole shadow radius can be calculated from the effective potential of test particles moving in black hole spacetime \cite{Hioki2008,Pugliese2011,Johannsen2013,Guo2020,Carroll,Hartle,Perlick2021,Raffaelli2021,Gan2021}. In recent years, other approaches on circular photon orbits (photon sphere, light rings) using topological and geometric techniques also emerged \cite{Virbhadra2001,Cunha2017,Cunha2020,Ghosh2021}.

In the present work, a pure geometric approach is developed to obtain the stable and unstable circular photon orbits, as well as black hole shadows. Our approach is implemented in the optical geometry of black hole spacetime. In this approach, the geodesic curvature and Gauss curvature in optical geometry turn out to be crucial quantities to determine the circular photon orbits. The stability of circular photon orbits is reflected by an elegant theorem in differential geometry and topology —— the Hadamard theorem. Furthermore, in this work, we also prove that the geometric approach developed in this work is completely equivalent to the conventional approach based on effective potential of test particles.

\textbf{Optical Geometry of Black Hole Spacetime:}
The optical geometry is a powerful tool to study the motions of photons (or other massless particles which travel along null geodesics) in gravitational field.  \cite{Abramowicz1988,Gibbons2008,Gibbons2009,Werner2012}.
For a four dimensional spacetime, its optical geometry can be constructed from the null constraint $d\tau^{2}=0$.
\begin{equation}
	\underbrace{d\tau^{2} = g_{\mu\nu}dx^{\mu}dx^{\nu}}_{\text{Spacetime Geometry}}
	\ \ \overset{d\tau^{2}=0}{\Longrightarrow} \ \ 
	\underbrace{dt^{2} = g^{\text{OP}}_{ij}dx^{i}dx^{j}}_{\text{Optical Geometry}}
	\label{aptical geometry}
\end{equation}
The properties of optical geometry are strongly depend on the symmetries of gravitational field and black hole spacetime. For a spherically symmetric black hole, its optical geometry gives a Riemannian manifold \cite{Abramowicz1988,Gibbons2008,Gibbons2009}. However, for a rotational black hole, the correspond optical geometry is a Randers-Finsler manifold \cite{Werner2012,Ono2017,Jusufi2018,Jusufi2018b}. Further, if we consider particle motions in the equatorial plane, a two dimensional manifold can be constructed from the optical geometry.
\begin{equation}
	\underbrace{dt^{2} = g^{\text{OP}}_{ij}dx^{i}dx^{j}}_{\text{Optical Geometry}}
	\ \ \overset{\theta=\pi/2}{\Longrightarrow} \ \ 
	\underbrace{dt^{2}=\tilde{g}^{\text{OP}}_{ij}dx^{i}dx^{j}}_{\text{Optical Geometry (Two Dimensional)}}
	\label{optical geometry2}
\end{equation}
In two dimensional manifold, a number of elegant and classical theorems in surface theory and differential geometry could provide useful tools to study the particle motions. In this work, our new geometric approach on circular photon orbits and black hole shadow is implemented in two dimensional optical geometry. A detailed description of optical geometry is given in supplemental material.

\textbf{Gauss Curvature and Geodesic Curvature:}
There are several key quantities which describe the geometric properties of optical geometry. It would be turned out that Gauss curvature and geodesic curvature in optical geometry play central roles in determining the circular photon orbits for black holes. The Gauss curvature is the intrinsic curvature of a two dimensional surface. The geodesic curvature is the curvature of one dimensional curves lived in two dimensional curved surface, which measures how far these curves are from being geodesics. If $\gamma$ is a geodesic curve in surface $S$, its geodesic curvature $\kappa_{g}(\gamma)$ automatically vanishes. Notably, both Gauss curvature and geodesic curvature only rely on the intrinsic metric of two dimensional surface $S$, regardless of the embedding of this surface into higher dimensional spacetime \cite{Berger}. More discussions on Gauss curvature and geodesic curvature can be found in the supplemental material.
 
\begin{figure}
	\includegraphics[width=0.5\textwidth]{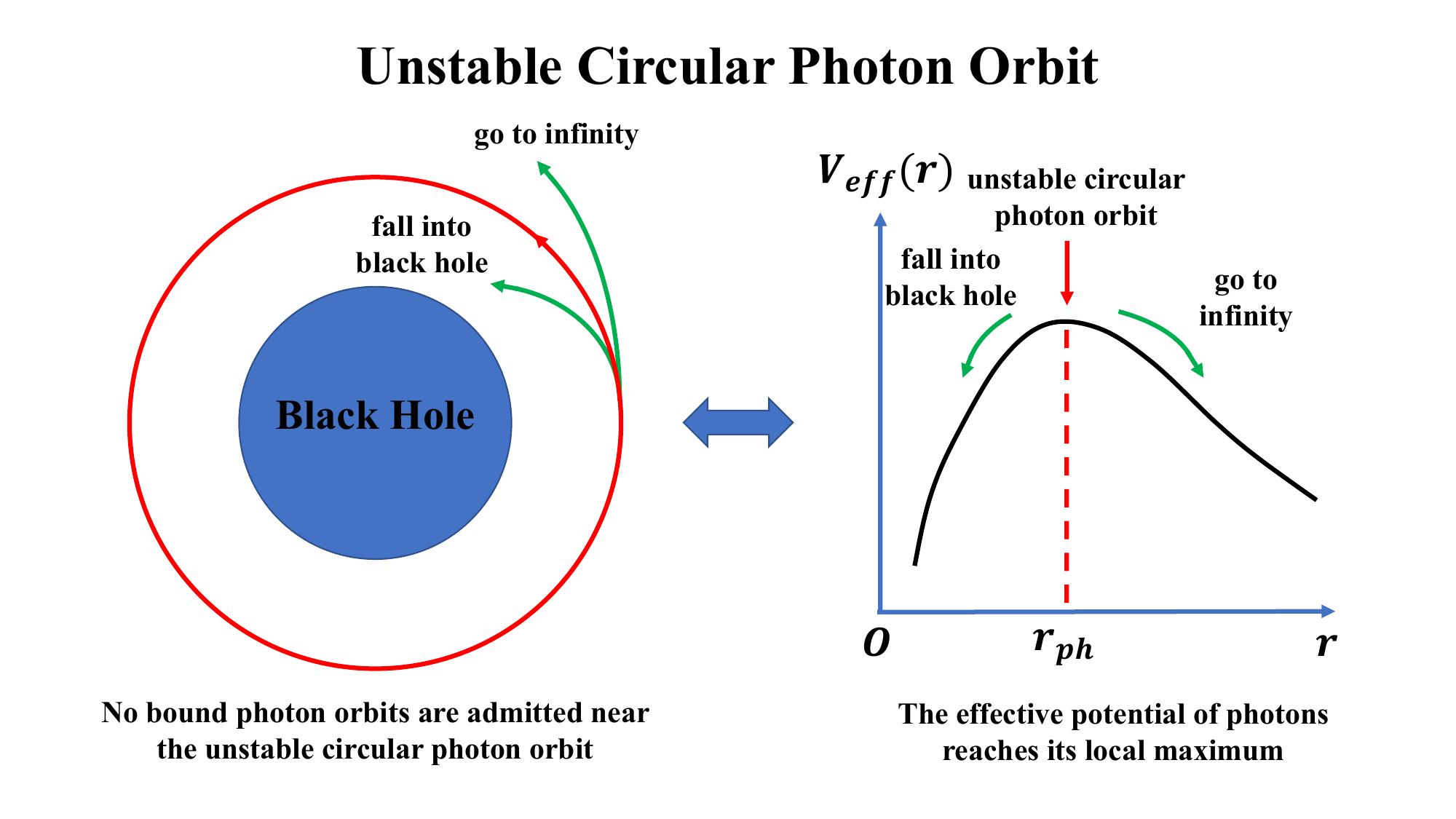}
	\includegraphics[width=0.5\textwidth]{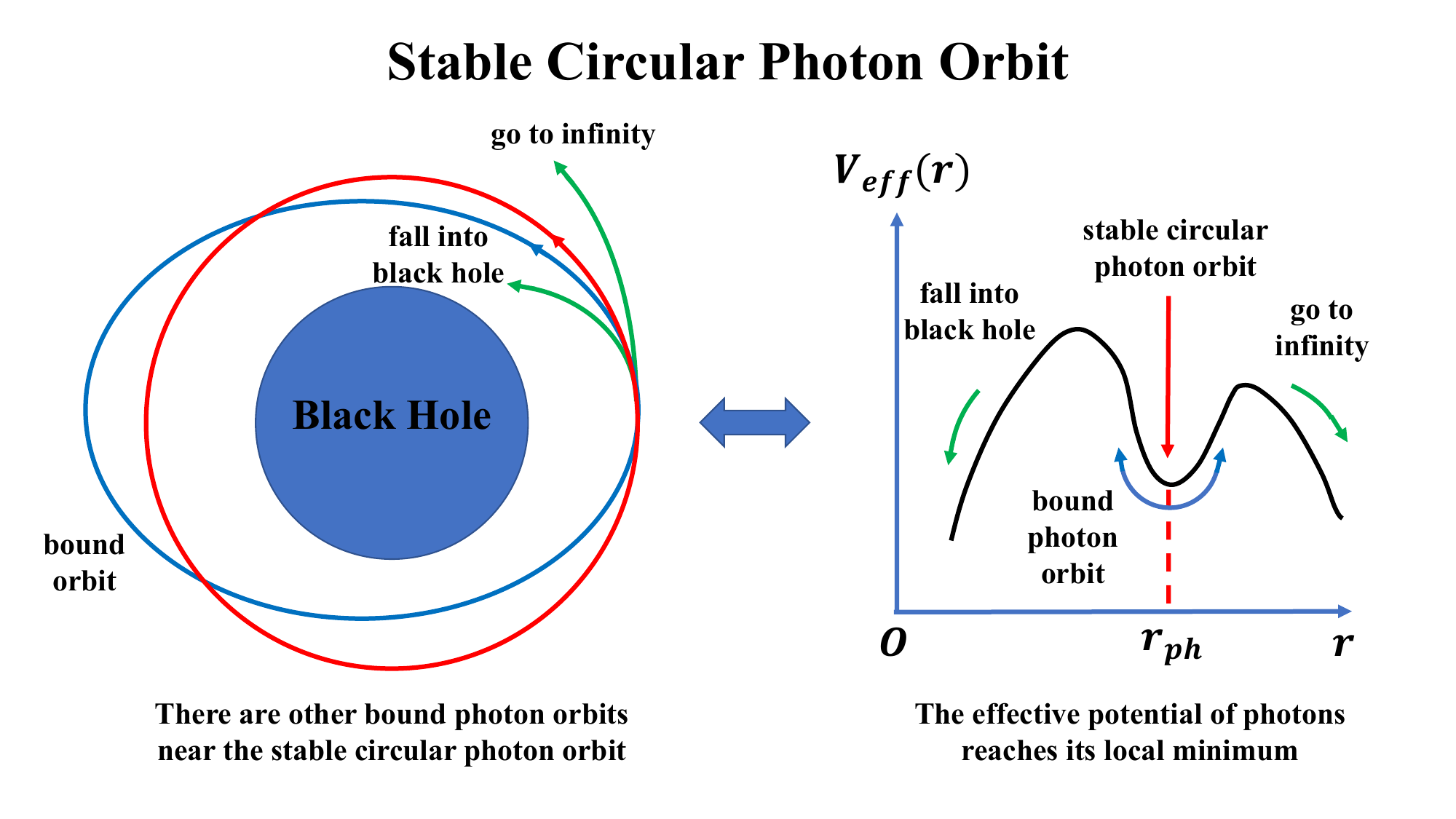}
	\caption{The schematic plot of the stable and unstable circular photon orbits near black holes. In this figure, we choose the spherically symmetric black hole as a typical example, and the region covered by blue is the interior of black hole.}
	\label{figure1}
\end{figure}

\textbf{Circular Photon Orbit:}
The circular photon orbits are important features of black holes. They are closely connected with particle motions, gravitational lensing and black hole shadows. The circular photon orbits are classified into two categories: the stable and unstable circular photon orbits. They may exhibit significantly different features, especially for the geodesics nearby. For unstable circular photon orbit, when photon beams have a departure from the circular orbit, they would either fall into black hole or move to infinity. No bound photon orbits are admitted near the unstable circular photon orbit. Conversely, there are many bound photon orbits near the stable circular photon orbit. These bound photon orbits may have different shapes. 
The stable and unstable circular photon orbits are illustrated in figure \ref{figure1}. 

In this work, a pure geometric approach is developed to obtain these circular photon orbits. We choose a class of stationary and spherically symmetric black holes to show our geometric approach. For spherically symmetric black hole with the metric
\begin{equation}
	d\tau^{2} = f(r)dt^{2} - \frac{1}{f(r)}dr^{2}
	-r^{2} (d\theta^{2}+\sin^{2}\theta d\phi^{2}) \label{spherical symmetric spacetime0}
\end{equation}
the optical geometry restricted in the equatorial plane $\theta=\pi/2$ gives 
\begin{equation}
	dt^{2}=\tilde{g}^{\text{OP}}_{ij}dx^{i}dx^{j}
	=\frac{1}{[f(r)]^{2}}dr^{2}+\frac{r^2}{f(r)}d\phi^{2}
	\label{optical metric11}
\end{equation}
Here, we only restrict to a subclass of static and spherically symmetric black hole such that the spacetime metric satisfies $g_{tt} \times g_{rr} = -1$ and $g_{\theta\theta} = r^{2}$. The cases of more general spherically symmetric and rotational (axisymmetric) black holes are left to incoming studies.

The circular photon orbits could be determined by Gauss curvature and geodesic curvature in optical geometry. Firstly, for circular photon orbit $r=r_{ph}$, which is a geodesic curve in optical geometry, its geodesic curvature vanishes naturally \cite{Carmo1976,ChernWH,footnote3}
\begin{eqnarray}
	\kappa_{g}(r=r_{ph}) & = & \frac{1}{2\sqrt{\tilde{g}^{\text{OP}}_{rr}}} 
	\frac{\partial \ log(\tilde{g}^{\text{OP}}_{\phi\phi})}{\partial r} \bigg|_{r=r_{ph}} \nonumber \\
	& = & \bigg[ \frac{f(r)}{r}-\frac{1}{2}\cdot\frac{\partial f(r)}{\partial r} \bigg]_{r=r_{ph}} 
	= 0 \label{geodesic cuurvature}
\end{eqnarray}
In this way, the radius of circular photon orbit is obtained. Here, we should emphasize that an important property of optical geometry is used in deriving this relation. The null geodesic curve $\gamma=\gamma(\tau)$ in the spacetime geometry $d\tau^{2}=g_{\mu\nu}dx^{\mu}dx^{\nu}$ maintains geodesic in the optical geometry $dt^{2}=g^{\text{OP}}_{ij}dx^{i}dx^{j}$ \cite{Gibbons2009}. Actually, it can be viewed as the generalization of Fermat's principle in curved stationary spacetime \cite{Gibbons2009,Werner2012,footnote1}.  

The following question is how to distinguish stable photon orbits from unstable photon orbits. The following Hadamard theorem in differential geometry would answer this question appropriately.
\begin{quote}
	\textbf{Hadamard Theorem:}
	For a two dimensional complete Riemannian manifold with  nonpositive Gauss curvature, there is only one geodesic curve from $p$ to $q$ belong to the same homotopy class, and this geodesic curve minimizes the length in this homotopy class \cite{Carmo1988,Jost2011}.
\end{quote} 
For spherically symmetric black hole, if we restrict the optical geometry in two dimensional equatorial plane, the Gauss curvature in the Hadamard theorem can be calculated through \cite{Carmo1976,ChernWH}
\begin{eqnarray}
	\mathcal{K} & = & -\frac{1}{\sqrt{\tilde{g}^{\text{OP}}}} 
		                 \bigg[
		                   \frac{\partial}{\partial \phi} \bigg( \frac{1}{\sqrt{\tilde{g}^{\text{OP}}_{\phi\phi}}} \frac{\partial\sqrt{\tilde{g}^{\text{OP}}_{rr}}}{\partial \phi}  \bigg)
		                   + \frac{\partial}{\partial r} \bigg( \frac{1}{\sqrt{\tilde{g}^{\text{OP}}_{rr}}} \frac{\partial\sqrt{\tilde{g}^{\text{OP}}_{\phi\phi}}}{\partial r}  \bigg)
		                 \bigg]  \nonumber
		               \\
		        & = & \frac{1}{2} f(r) \cdot \frac{d^{2}f(r)}{dr^{2}}
		              - \bigg[ \frac{1}{2} \cdot \frac{df(r)}{dr} \bigg]^{2} \label{Gauss curvature}
\end{eqnarray}
where $\tilde{g}^{\text{OP}}=det(\tilde{g}^{\text{OP}}_{ij})$ is determinant of optical metric.

\begin{figure}
	\includegraphics[width=0.435\textwidth]{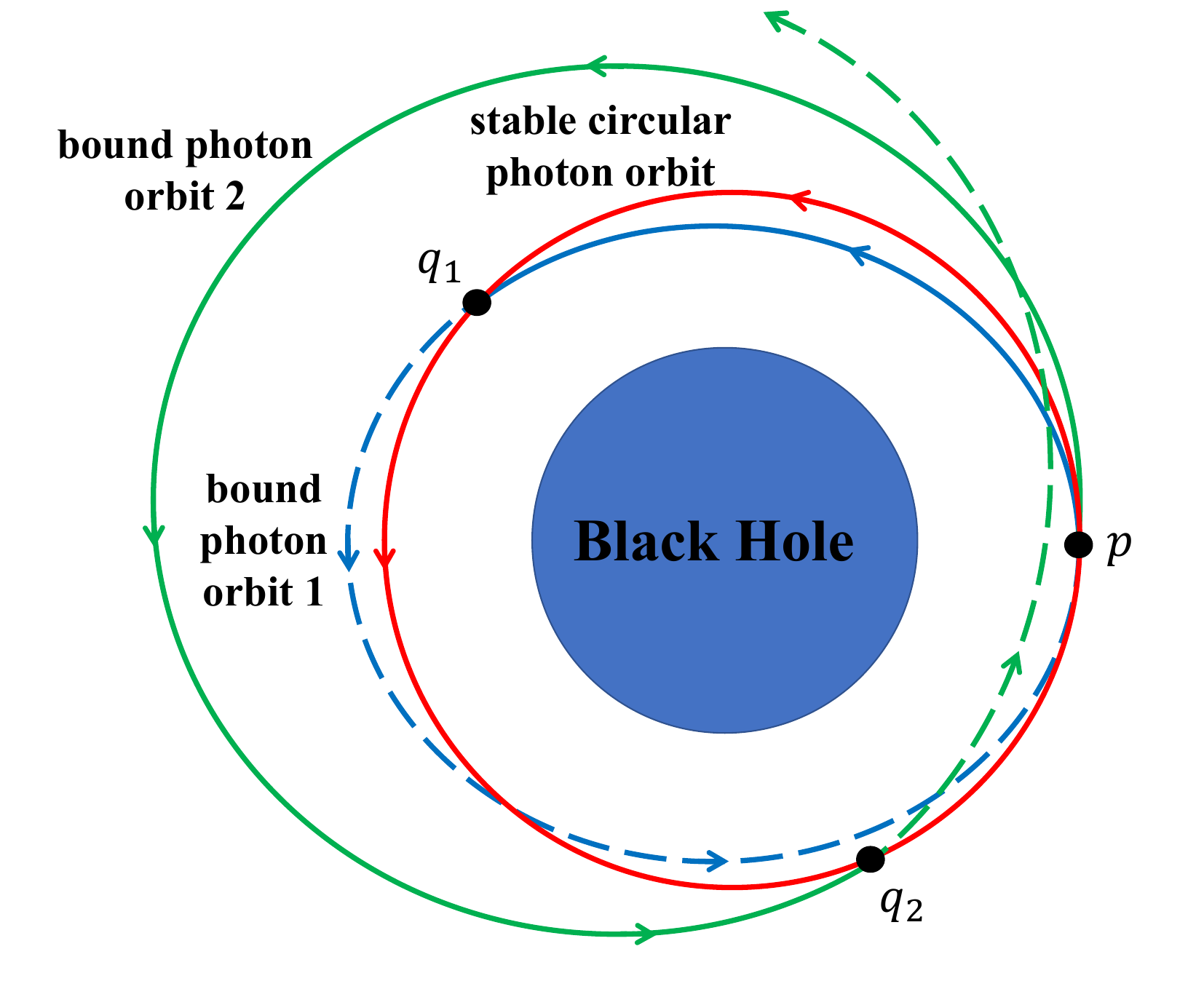}
	\caption{This figure shows several possible bound photon orbits near the stable circular photon orbits. These bound photon orbits could be obtained by perturbation of circular photon orbit at point $p$, and they may have different shapes. The bound photon orbit $1$ labeled in blue is a closed orbit, and the bound photon orbit $2$ labeled in green is an unclosed orbit (just like the Mercury procession around the Sun). In the equatorial plane of optical geometry, we can find another point $q$ (we denote the $q_{i}$ for bound orbit $i=1,2$). When we considering photon orbits starting from $p$ ending with $q$, there are at least two different photon orbits (one is the stable circular photon orbit, and the other is a bound photon orbit) that can be continuously deformed to each other. Therefore, in the equatorial plane of optical geometry, there are two different geodesic curves from $p$ to $q$ belong to the same homotopy class.}
	\label{figure2}
\end{figure} 

For stable circular photon orbit, there are bound photon orbits nearby, which may have different shapes. In the equatorial plane of optical geometry, we can find two points $p$ and $q$ such that there are at least two geodesic curves (one is the stable photon orbit, the other is a bound photon orbit) belong to the same homotopy class. Figure \ref{figure2} illustrates two possible bound photon orbits and the corresponding choice of points $p$ and $q$. In such cases, the Gauss curvature of optical geometry should be positive, otherwise it would violate the Hadamard theorem. On the contrary, for unstable circular photon orbit, no bound photon orbits homotopic to this circular photon orbit exist. Then unstable circular photon orbit itself forms the whole homotopy class, which correspond to the negative Gauss curvature in the Hadamard theorem (we assume the Gauss curvature of optical geometry is nonzero, otherwise the black hole spacetime would by flat). Based on the above discussions, we obtain the following criterion to determine the stable and unstable circular photon orbits \cite{footnote2}. 
\begin{eqnarray}
	\mathcal{K} < 0 & \Rightarrow & \text{The circular photon orbit $r=r_{ph}$ is unstable}  \nonumber \\
	\mathcal{K} > 0 & \Rightarrow & \text{The circular photon orbit $r=r_{ph}$ is stable} \nonumber
\end{eqnarray}

\begin{figure}
	\includegraphics[width=0.5\textwidth]{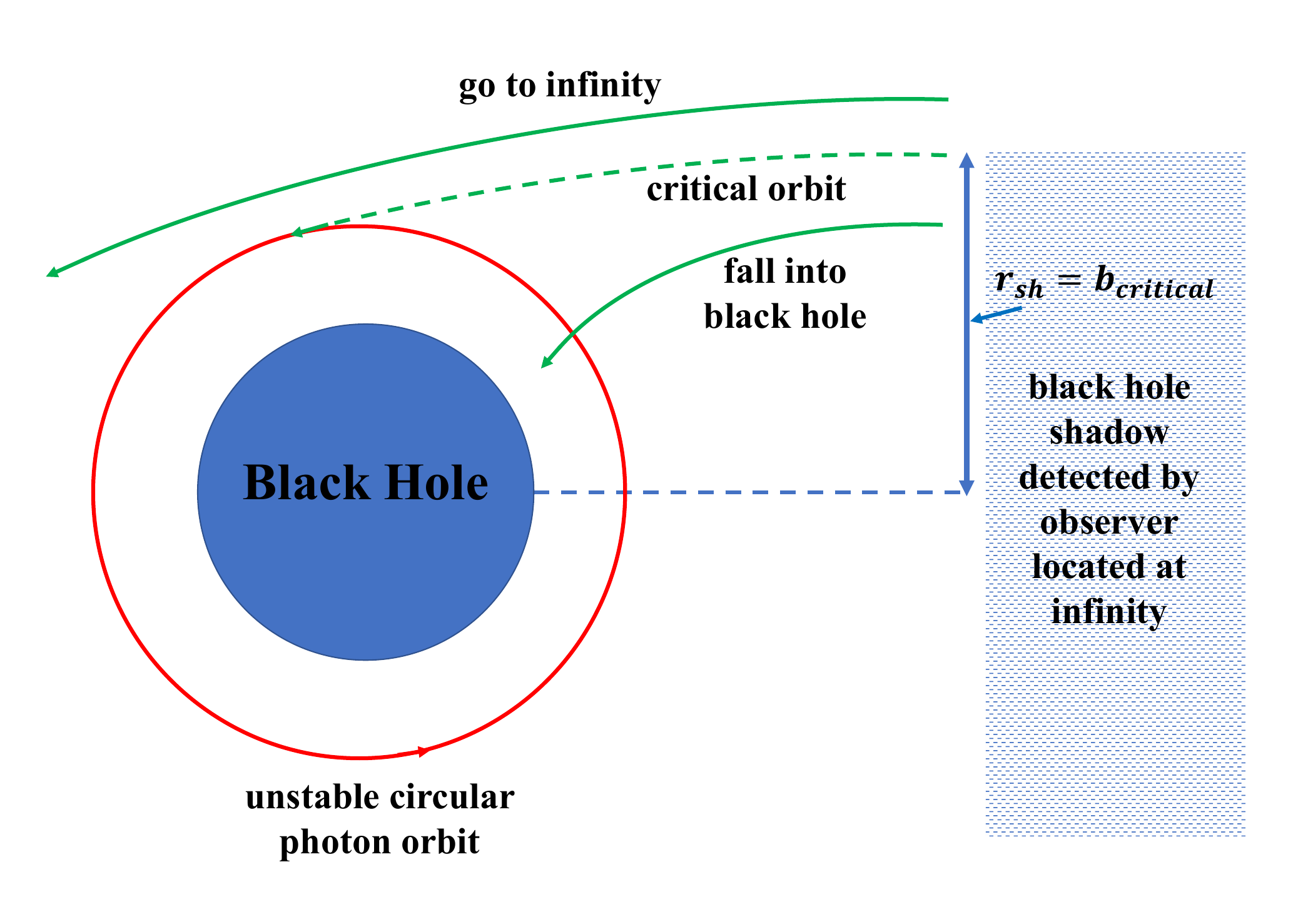}
	\caption{The black hole shadow and unstable circular photon orbit. The light beam emitted from infinity with impact parameter $b=b_{\text{critical}}$ would reach the unstable circular photon orbit exactly. The light beam emitted with impact parameter $b<b_{\text{critical}}$ would fall into the black hole.}
	\label{figure3}
\end{figure}

\textbf{Black Hole Shadow:} 
Black hole shadow is the dark silhouette of black hole image in a bright background. The size and shape of black hole shadow depends not only on black hole parameters, but also on the position of observers \cite{Perlick2021}. In this work, we concentrate on the idealized situation where observer is located at infinity, and there are no light sources between the observer and black hole. In this case, the radius of black hole shadow detected by observer is just the critical value of impact parameter $b_{\text{critical}}$. In the gravitational field, light beams emitted from infinity with impact parameter $b=b_{\text{critical}}$ would reach the unstable circular photon orbit exactly, as illustrated in figure \ref{figure3}

\begin{table*}
	\caption{The circular photon orbit and black hole shadow radius detected by observer at infinity. This table summarize the results for Schwarzschild black hole and Reissner-Nordstr\"om black hole. All quantities calculated using our new geometric approach agree with the results obtained using conventional approach based on effective potential. \footnote{Note that optical geometry is defined outside the black hole horizon, which is $r>2M$ for Schwarzschild black hole and $r>M+\sqrt{M^{2}-Q^{2}}$ for Reissner-Nordstr\"om black hole.}}
	\label{table1}
	\vspace{2mm}
	\begin{ruledtabular}
		\begin{tabular}{lcccc}
			& Type of Black Hole & Schwarzschild Black Hole & Reissner-Nordstr\"om Black Hole &
			\\
			\hline
			& Metric & $f(r)=1-2M/r$ & $f(r)=1-2M/r+Q^{2}/r^{2}$ 
			\\
			& Geodesic Curvature & $\kappa_{g} = 1/r-3M/r^{2}$ & $\kappa_{g} = 1/r-3M/r^{2}+2Q^{2}/r^{3} $ &
			\\
			& Circular Photon Orbit & $r_{ph}=3M$ & $r_{ph}=(3M+\sqrt{9M^{2}-8Q^{2}})/2$ &
			\\
			& Gauss Curvature & $\mathcal{K}=-2M/r^{3}+3M^{2}/r^{4}<0$ & $\mathcal{K}=-2M/r^{3}+3(M^{2}+Q^{2})/r^{4}-6MQ^{2}/r^{5}+2Q^{4}/r^{6}<0$&
			\\
			& & (outside black hole horizon, $r>2M$) & (outside black hole horizon, $r>M+\sqrt{M^{2}-Q^{2}}$) &
			\\
			& Stability & $r_{ph}=3M$ is unstable circular orbit. &  $r_{ph}=(3M+\sqrt{9M^{2}-8Q^{2}})/2$ is unstable circular orbit. &
			\\
			& Black Hole Shadow & $r_{sh}=b_{\text{critical}}=3\sqrt{3}M$ & $r_{sh}=b_{\text{critical}}=\frac{(3M+\sqrt{9M^{2}-8Q^{2}})^{2}}{\sqrt{8(3M^{2}-2Q^{2}+M\sqrt{9M^{2}-8Q^{2}})}}$ &
		\end{tabular}
	\end{ruledtabular}
\end{table*}

Following the conventional definition, the impact parameter can be expressed as \cite{Hartle}
\begin{equation}
	b \equiv \bigg|\frac{L}{E}\bigg|
	=   \bigg|
	\frac{r^{2}\sin^{2}\theta \cdot d\phi/d\lambda}{f(r)\cdot dt/d\lambda}
	\bigg|
\end{equation}
where $E$ and $L$ is the conserved energy and angular momentum per unit mass
\begin{subequations}
	\begin{eqnarray}
		E = f(r)\cdot\frac{dt}{d\lambda}  
		\\
		L = r^{2}\sin^{2}\theta\cdot\frac{d\phi}{d\lambda}
	\end{eqnarray}
\end{subequations}
Since the impact parameter $b \equiv L/E$ 
is a conserved quantity along the geodesics, we can calculate the critical impact parameter $b_{\text{critical}}$ at any specific point along the photon trajectory. A simple and convenient choice are the points in the unstable circular photon orbit $r=r_{\text{unstable}}$ 
\begin{equation}
	b_{\text{critical}} = \bigg| \frac{L}{E} \bigg|_{r=r_{\text{unstable}}}
	= \bigg| \frac{r^{2}}{f(r)} \cdot \frac{d\phi/d\lambda}{dt/d\lambda} \bigg|_{r=r_{\text{unstable}}}
\end{equation}
where $\theta=\pi/2$ has been used in the equatorial plane.

The radius of black hole shadow detected by observer at infinity 
can be calculated in the optical geometry. In the spacetime geometry, photon orbits are along lightlike / null geodesics, and their tangent vector satisfies $T \cdot T = g_{\mu\nu} \frac{dx^{\mu}}{d\lambda} \frac{dx^{\nu}}{d\lambda} = 0$, with $\lambda$ to be any affine parameter of null geodesics. However, when transformed into the optical geometry, the photon orbits become spatial geodesics with vanishing geodesic curvature $\kappa_{g}=0$, as we have emphasized previously \cite{footnote3}. The stationary time coordinate $t$ in spacetime geometry exactly reduces to the arc-length parameter / spatial distance parameter in optical geometry. 
In the optical geometry, the tangent vector of arbitrary photon orbits $T^{\text{OP}}=d/dt$ with respect to arc-length parameter $t$ becomes a unit vector ($|T^{\text{OP}}|=1$). Here, we restrict the optical geometry in the equatorial plane, the tangent vector of photon orbits reduce to $T^{\text{OP}}=(\frac{dr}{dt},\frac{d\phi}{dt})$. Furthermore, for unstable circular photon orbit, 
we have the result
\begin{equation}
	\begin{split}
	& r=r_{\text{unstable}}=\text{constant} 
	\ \Rightarrow \ 
	\frac{dr}{dt} \bigg|_{r=r_{\text{unstable}}} = 0 
	\\
    \Rightarrow & 
    |T^{\text{OP}} \cdot T^{\text{OP}}| = \bigg| \tilde{g}^{\text{OP}}_{\phi\phi} \cdot  \frac{d\phi}{dt} \cdot \frac{d\phi}{dt} \bigg| 
    \\
    & \ \ \ \ \ \ \ \ \ \ \ \ \ \ \ \ 
    = \bigg| \frac{r^{2}}{f(r)} \cdot \frac{d\phi}{dt} \cdot \frac{d\phi}{dt} \bigg|_{r=r_{\text{unstable}}}
    = 1 
    \\
    \Rightarrow & r_{sh} = b_{\text{critical}} 
    = \bigg| \frac{r^{2}}{f(r)} \cdot \frac{d\phi/d\lambda}{dt/d\lambda} \bigg|_{r=r_{\text{unstable}}} 
    = \sqrt{\frac{r_{\text{unstable}}^{2}}{f(r_{\text{unstable}})}} 
    \end{split}
\end{equation}
This is the analytical expression for black hole shadow radius detected by observer located at infinity, and it is consistent with previous results \cite{Guo2020,Perlick2021}. Based on above procedures, the shadow radius for arbitrary spherically symmetric black hole with metric from in equation (\ref{spherical symmetric spacetime0}) can be calculated. In the present work, we choose two typical examples in general relativity —— Schwarzschild black hole and Reissner-Nordstr\"om black hole. The circular photon orbits and black hole shadow radius for these black holes are summarized in table \ref{table1}.

\textbf{Equivalence between Our Geometric Approach and Conventional Approach:}
The geometric approach developed in this work is completely equivalent to the conventional approach based on effective potentials of test particles moving in the gravitational field. In conventional approach, the circular photon orbit can be solved by analyzing the extreme points of effective potential $V_{\text{eff}}(r)$. Particularly, the unstable circular photon orbit $r=r_{\text{unstable}}$ corresponds to local maximum of effective potential, and the stable circular photon orbit $r=r_{\text{stable}}$ is the local minimum of effective potential. 

\begin{table*}
	\caption{The features of our geometric approach developed in this work and the conventional approach.}
	\label{table2}
	\vspace{2mm}
	\begin{ruledtabular}
		\begin{tabular}{lcccc}
			& Approach                 & Our Geometric Approach & Conventional Approach &
			\\
			\hline
			& Geometry               & Optical Geometry of Spacetime  & Spacetime Geometry &
			\\
			&                        & (Riemannian / Randers-Finsler Geometry) \footnote{This table summarizes the results on spherically symmetric black holes, in which the optical geometry is Riemannian manifold. The rotational symmetric black holes, whose optical geometry is Randers-Finsler manifold, is left to on-going work.} & (Lorentz Geometry) &
			\\
			\hline
			& Key Quantities         & Gauss Curvature $\mathcal{K}(r)$ & Effective Potential $V_{\text{eff}}(r)$ &
			\\
			&                        & Geodesic Curvature $\kappa_{g}(r)$ &       
			\\
			\hline
			& Photon Orbit           & Conditions  & Conditions &
			\\
			\hline
			& circular photon orbit  & zero geodesic curvature & extreme point of effective potential &
			\\
			&                        & $\kappa_{g}(r)=0$ & $\frac{dV_{\text{eff}}(r)}{dr}=0$ &
			\\
			& unstable circular photon orbit \  & zero geodesic curvature and negative Gauss curvature \ \    & local maximum of effective potential &
			\\
			&                                   & $\kappa_{g}(r)=0$ and $\mathcal{K}(r)<0$ & $\frac{dV_{\text{eff}}(r)}{dr}=0$ and $\frac{d^{2}V_{\text{eff}}(r)}{dr^{2}}<0$ &
			\\
			& stable circular photon orbit \    & zero geodesic curvature and positive Gauss curvature \ \    & local minimum of effective potential &
			\\
			&                                   & $\kappa_{g}(r)=0$ and $\mathcal{K}(r)>0$ & $\frac{dV_{\text{eff}}(r)}{dr}=0$ and $\frac{d^{2}V_{\text{eff}}(r)}{dr^{2}}>0$ &
			\\
		\end{tabular}
	\end{ruledtabular}
\end{table*}

We now demonstrate the equivalence between these two approaches. In spherically symmetric black hole spacetime, if we restrict test particles moving in the equatorial plane $\theta=\pi/2$, the equation of motion eventually reduces to \cite{Carroll}
\begin{equation}
	\frac{1}{2}\cdot\bigg(\frac{dr}{d\lambda}\bigg)^{2}+V_{\text{eff}}(r)
	= \frac{1}{2} \cdot E^{2}
\end{equation}
where effective potential of test particles is defined by
\begin{equation}
	V_{\text{eff}}(r) \equiv 
	\frac{f(r)}{2} \cdot \bigg(\frac{L^{2}}{r^{2}}+\epsilon\bigg)
\end{equation}
Here, for massless particles $\epsilon=0$, and for massive particles $\epsilon=1$. 
The circular photon orbit $r=r_{ph}$ correspond to the extreme point of effective potential \cite{Carroll,Hartle}
\begin{eqnarray}
	& & \frac{dV_{\text{eff}}(r)}{dr} \bigg|_{r=r_{ph}} = 0 \nonumber
	\\ 
	& \Rightarrow  &
	\bigg[
	\frac{L^{2}}{2r^{2}}\cdot\frac{df(r)}{dr}
	-\frac{L^{2}\cdot f(r)}{r^{3}}
	\bigg]_{r=r_{ph}} 
	= 0 \label{extreme point equation0}
\end{eqnarray}
where we have used $\epsilon=0$ for massless photons. Note that the  angular momentum $L$ is a conserved quantity along photon orbit, which is independent of $r$. Comparing equation (\ref{extreme point equation0}) with equation (\ref{geodesic cuurvature}), it is clearly demonstrated that they are equivalent to each other. When $r=r_{ph}$ is the extreme point of effective potential $V_{\text{eff}}(r)$, the geodesic curvature along the circle $r=r_{ph}$ vanishes precisely, which makes it to be the circular photon orbit for spherically symmetric black holes.

Then we analyze the stability of circular photon orbits. For unstable circular photon orbit $r=r_{\text{unstable}}$, the effective potential should reach its local maximum \cite{Hartle,Raffaelli2021}
\begin{equation}
	\begin{split}
	& \frac{d^{2}V_{\text{eff}}(r)}{dr^{2}} \bigg|_{r=r_{\text{unstable}}} < 0 
	\\ 
	\Rightarrow &  
	\bigg[
	  \frac{L^{2}}{2r^{2}}\cdot\frac{d^{2}f(r)}{dr^{2}}
	  -\frac{2L^{2}}{r^{3}}\cdot\frac{df(r)}{dr}
	  +\frac{3L^{2}\cdot f(r)}{r^{4}} 
	\bigg]_{r=r_{\text{unstable}}} 
	<0 
	\\
	\Rightarrow &  
	\bigg[
	  \frac{L^{2}}{2r^{2}}\cdot\frac{d^{2}f(r)}{dr^{2}}
	  -\frac{L^{2}\cdot f(r)}{r^{4}}
	\bigg]_{r=r_{\text{unstable}}} 
    < 0 
    \\ 
    \Rightarrow & 
    \bigg[
      \frac{f(r)}{2}\cdot\frac{d^{2}f(r)}{dr^{2}}
      -\frac{f(r) \cdot f(r)}{r^{2}}
    \bigg]_{r=r_{\text{unstable}}} 
    < 0 
    \\ 
    \Rightarrow & 
    \bigg[
      \frac{f(r)}{2}\cdot\frac{d^{2}f(r)}{dr^{2}}
      -\bigg(\frac{1}{2}\frac{df(r)}{dr}\bigg)^{2}
    \bigg]_{r=r_{\text{unstable}}} 
    < 0 
    \end{split}
    \label{maximum point equation0}
\end{equation}
In the derivation, we have used the relation $df(r)/dr=2f(r)/r$ in equation (\ref{extreme point equation0}) for unstable circular photon orbit 
repeatedly. Recall the expression for Gauss curvature in equation (\ref{Gauss curvature}), the last line of this inequality implies that Gauss curvature must be negative ($\mathcal{K}<0$). On the contrary, the stable photon orbit, which correspond to the local minimum of effective potential, would imply the Gauss curvature to be positive ($\mathcal{K}>0$). 
In this way, the equivalence between our geometric approach developed in this work and the conventional approach is demonstrated. The features of these two approaches and their equivalence are summarized in table \ref{table2}.

\textbf{Summary and Prospects:}
In this work, a pure geometric approach is developed to calculate the circular photon orbit and black hole shadow radius. This approach is quite simple and general, regardless of particular metric forms of black hole spacetime. The Gauss curvature, geodesic curvature in optical geometry and Hadamard theorem in differential geometry offer a new pathway to calculate the circular photon orbits and black hole shadows. This approach indicates that the optical geometry may give us profound insights on black hole properties, and it is worthy of extensive investigations. Furthermore, we demonstrate that our approach is completely equivalent to the conventional approach based on effective potentials of test particles.

There are several possible extensions of our geometric approach in the near future. Firstly, the similar algorithm can be apply to rotational black holes, whose optical geometry is a Randers-Finsler manifold. Secondly, the approach developed in this work, which is given for massless photons, can also be generated to the cases of massive particles. For circular orbits of massive particles, the utilization of Gauss curvature, geodesic curvature and Hadamard theorem may take place in the Jacobi geometry of black hole spacetime, rather than the optical geometry. The Jacobi geometry can be constructed from the action principle and constrained canonical momenta of massive test particles \cite{Gibbons2016,Gibbons2019}.

\begin{acknowledgments}
	\textbf{Acknowledgments:} This work was supported by the National Natural Science Foundation of China (Grants No. 11871126) and the Scientific Research Foundation of Chongqing University of Technology (Grants No. 2020ZDZ027).
\end{acknowledgments}






\begin{thebibliography}{99}
\bibitem{footnote2}
	Although black hole spacetime has singularities, its optical geometry is defied outside the horizons, so it is usually a geodesic complete manifold. Since time coordinate $t$ in spacetime geometry becomes arc length parameter in optical geometry, the endless of time coordinate $t$ outside horizons would imply the geodesic completeness of optical geometry.
\end{thebibliography}

\end{document}



\title{Supplemental Material: \\ Geometric approach to circular photon orbits and black hole shadows}


\author{Chen-Kai Qiao}
\email{chenkaiqiao@cqut.edu.cn}
\affiliation{College of Science, Chongqing University of Technology, Banan, Chongqing, 400054, China}

\author{Ming Li}
\email{mingli@cqut.edu.cn}
\affiliation{Mathematical Science Research Center, Chongqing University of Technology, Banan, Chongqing, 400054, China}


\date{\today}

\begin{abstract}
This supplemental material gives a detailed description on the optical geometry of black hole spacetime, as well as the Gauss curvature and geodesic curvature used in our geometric approach. The first part discusses the optical geometry of spherically symmetric and axi-symmetric black holes. The second part introduces the Gauss curvature and geodesic curvature in surface theory and differential geometry. 
\end{abstract}




\maketitle

\section{1. Optical Geometry of Black Hole Spacetime \label{supplementary1}}

This section gives an introduction of the optical geometry of black hole spacetime. The optical geometry can be obtained from spacetime geometry $d\tau^{2}=g_{\mu\nu}dx^{\mu}dx^{\nu}$ by imposing the null constraint $d\tau^{2}=0$. The optical geometry attracted a number of interests in recent years, and many aspects of black hole physics can be analyzed and studied using optical geometry \cite{Abramowicz1988,Gibbons2008,Gibbons2009}. Particularly, for stationary black hole spacetime, the photon orbits (which are along lightlike / null geodesics in four dimensional spacetime / Lorentz manifold), become spatial geodesics and minimize the stationary time $t$ when transformed into optical geometry. The stationary time coordinate $t$ plays the role of arc length parameter / spatial distance parameter in optical geometry, and $\delta[\int_{\gamma}dt]=0$ is satisfied along the photon orbits. This conclusion can be viewed as the generalization of the renowned Fermat's principle in curved manifold  \cite{Gibbons2009,Werner2012}, as we have discussed in the main text. In our geometric approach, both the circular photon orbits and black hole shadow radius are analyzed and calculated in the equatorial plane of optical geometry. 

The properties of optical geometry strongly depend on symmetries of gravitational field and black hole spacetime. For a spherically symmetric black hole, its optical geometry gives a Riemannian manifold \cite{Abramowicz1988,Gibbons2008,Gibbons2009}. To clarify this point, consider a spherically symmetric black hole spacetime with the metric
\begin{equation}
	d\tau^{2} = g_{tt}dt^{2}-g_{rr}dr^{2}-g_{\theta\theta}d\theta^{2}-g_{\phi\phi}d\phi^{2}
\end{equation} 
its optical geometry is obtained from the null constraint $d\tau^{2}=0$, which eventually gives a three dimensional Riemannian manifold
\begin{equation}
	dt^{2} = g_{ij}^{\text{OP}}dx^{i}dx^{j}
	       = \frac{g_{rr}}{g_{tt}} \cdot dr^{2} 
	         + \frac{g_{\theta\theta}}{g_{tt}} \cdot d\theta^{2} 
	         + \frac{g_{\phi\phi}}{g_{tt}} \cdot d\phi^{2}
\end{equation}
When studying the circular photon orbits, one can always restricts this optical geometry to the equatorial plane $\theta=\pi/2$ without loss of generality. Besides, for static and spherically symmetric black holes, another equivalent way to define optical geometry is from the conformal transformation of spacetime and extracting its spatial part \cite{Gibbons2009}.

For rotational / axi-symmetric black holes, the correspond optical geometry is a Randers-Finsler manifold \cite{Werner2012,Ono2017,Jusufi2018}. In these cases, we should consider the standard rotational black hole metric
\begin{equation}
	d\tau^{2} = g_{tt}dt^{2}-2g_{t\phi}dtd\phi-g_{rr}dr^{2}-g_{\theta\theta}d\theta^{2}-g_{\phi\phi}d\phi^{2}
\end{equation} 
the optical geometry can be obtained in a similar way by imposing the null constraint $d\tau^{2}=0$. The arc length / spatial distance in optical geometry eventually reduces to
\begin{equation}
	dt = \sqrt{\frac{g_{rr}}{g_{tt}} \cdot dr^{2} + \frac{g_{\theta\theta}}{g_{tt}} \cdot d\theta^{2} + \frac{g_{t\phi}^{2}+g_{tt}g_{\phi\phi}}{g_{tt}^{2}} \cdot d\phi^{2}} + \frac{g_{t\phi}}{g_{tt}}\cdot d\phi
\end{equation}
This gives a class of Renders-Finsler manifolds. The Renders-Finsler geometry is the extension of Riemannian geometry \cite{ChernSS}, in which the arc length / spatial distance can be separated into two parts
\begin{equation}
	dt = \sqrt{\alpha_{ij}dx^{i}dx^{j}} + \beta_{i}dx^{i} \label{Renders geometry}
\end{equation}
Here, the first part $\alpha_{ij}$ is a Riemannian metric, and the second part $\beta=\beta_{i}dx^{i}$ is an one-form which measures the departure of this Renders-Finsler geometry from the Riemannian geometry $dt^{2}=\alpha_{ij}dx^{i}dx^{j}$. The above Renders-Finsler geometry reduces to Riemannian geometry if and only if $\beta=0$.

In this work, we only duel with a class of spherically symmetric black holes, whose optical geometry is Riemannian manifold. Therefore, in the present work, both Gauss curvature and geodesic curvature are calculated using formulas in Riemannian geometry. The investigations on circular photon orbits and black hole shadows for rotational black holes, whose optical geometry gives a Renders-Finsler manifold, are left for incoming studies. 

\section{2. Gauss Curvature and Geodesic Curvature in Two Dimensional Surface \label{supplementary2}}

The section gives an introduction of the Gauss curvature and geodesic curvature in surface theory and differential geometry. In our geometric approach, both the circular photon orbits and black hole shadow radius are calculated and analyzed in the optical geometry of black hole spacetime, rather than in four dimensional spacetime (Lorentz manifold). 
In this supplemental material, we mainly focus on the Gauss curvature and geodesic curvature defined in two dimensional optical geometry $dt^{2}=\tilde{g}^{\text{OP}}_{ij}dx^{i}dx^{j}$ (the optical geometry restricted in equatorial plane $\theta=\pi/2$), which is always a two dimensional Riemanian manifold for spherically symmetric black holes. 

The Gauss curvature is the intrinsic curvature of a two dimensional surface $S$, which measures how far this surface is from being flat intrinsically. The geodesic curvature is the curvature of a curve $\gamma(s)$ lived in this surface, which measures how far this curve is from being a geodesic in surface $S$. There are many equivalent ways to define geodesic curvature and Gauss curvature. For instructive reasons, we embed the two dimensional curved surface $S$ into higher dimensional Euclidean space. However, both Gauss curvature and geodesic curvature are intrinsic properties of two dimensional surface $S$, irrespective of the embedding of surface $S$ into higher dimensional space. 


\begin{figure}
	\includegraphics[width=0.625\textwidth]{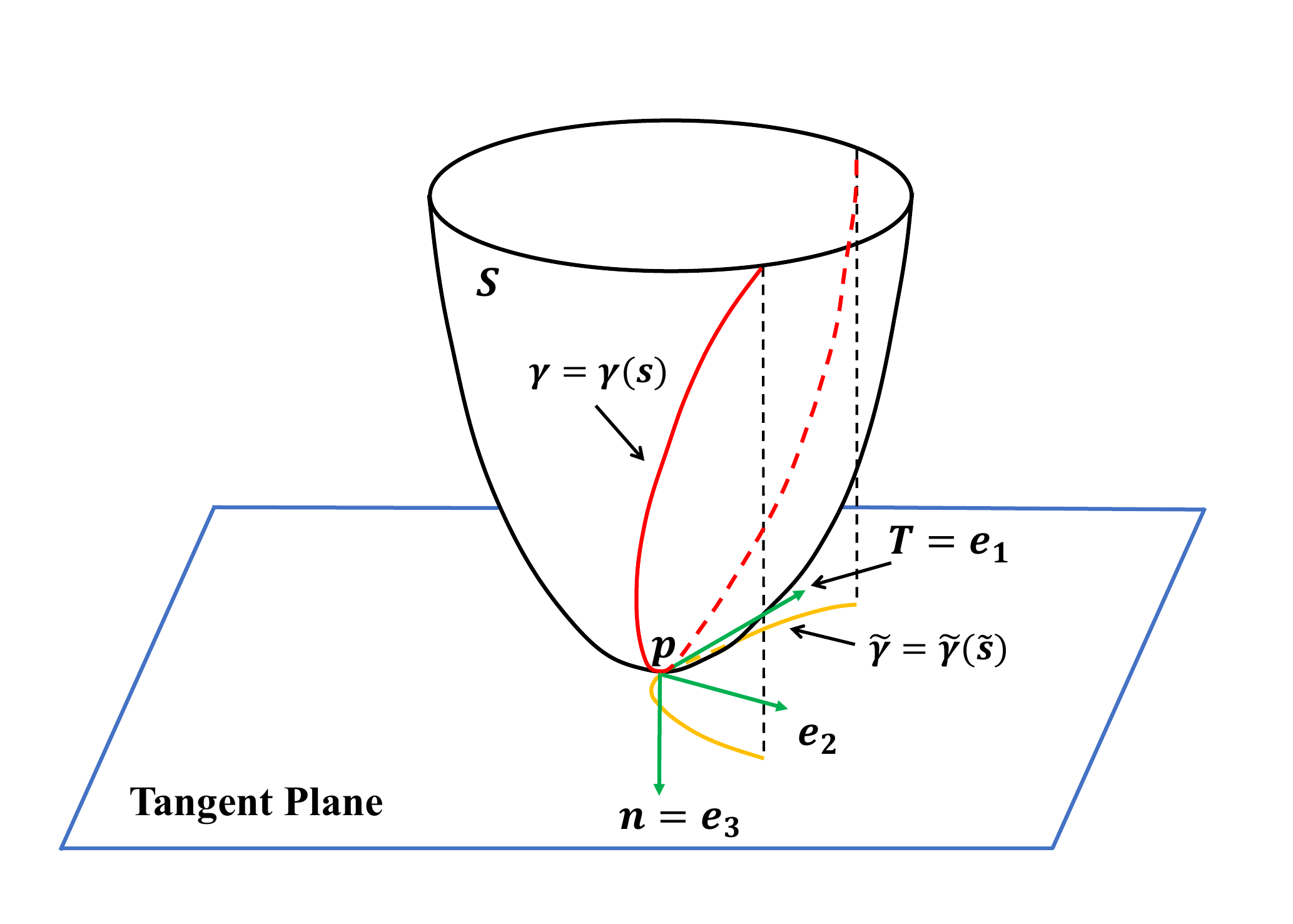}
	\caption{Illustration of the embedding of surface $S$ in three dimensional Euclidean space. In this figure, $S$ is a two dimensional curved surface, $\gamma=\gamma(s)$ is a curve lived in surface $S$, and $p$ is an arbitrary point along this curve. The $\{p; \boldsymbol{e_{1}},\boldsymbol{e_{2}},\boldsymbol{e_{3}}\}$ are frame fields in three dimensional Euclidean space such that $\boldsymbol{e_{1}}=\boldsymbol{T}$ is the unit tangent vector of curve $\gamma=\gamma(s)$ and $\boldsymbol{e_{3}}=\boldsymbol{n}$ is the unit normal vector of surface $S$.  The tangent plane of curved surface $S$ at point $p$ are generated by frames $\boldsymbol{e_{1}}$ and $\boldsymbol{e_{2}}$. The curve $\widetilde{\gamma}=\widetilde{\gamma}(\widetilde{s})$ is the projection of $\gamma=\gamma(s)$ on tangent plane of surface $S$ at point $p$.}
	\label{figure3}
\end{figure}

Consider a two dimensional curved surface $S$ embedded into three dimensional Euclidean space, which is illustrated in figure \ref{figure3}. For a curve $\gamma=\gamma(s)$ lived in surface $S$ with arc length parameter / spatial distance parameter $s$, its unit tangent vector in three dimensional space is $\boldsymbol{T}=d\boldsymbol{r}/ds$. The total curvature of $\gamma(s)$, which is the curvature of this curve viewed in three dimensional Euclidean space, is defined to be the module of derivative $d\boldsymbol{T}/ds$ \cite{Carmo1976,ChernWH}
\begin{equation}
	\kappa = \bigg| \frac{d\boldsymbol{T}}{ds} \bigg| 
	= \bigg| \frac{d^{2}\boldsymbol{r}(s)}{ds^{2}} \bigg|
\end{equation} 
For any point $p$ along this curve, 
we can assign the frame fields $\{p; \boldsymbol{e_{1}},\boldsymbol{e_{2}},\boldsymbol{e_{3}}\}$ in three dimensional space such that $\boldsymbol{e_{1}}=\boldsymbol{T}$ is the unit tangent vector of curve $\gamma=\gamma(s)$ and $\boldsymbol{e_{3}}=\boldsymbol{n}$ is the unit normal vector of surface $S$. The tangent plane of curved surface $S$ at point $p$ are generated by frames $\boldsymbol{e_{1}}$ and $\boldsymbol{e_{2}}$.  
In this case, the geodesic curvature of curve $\gamma=\gamma(s)$ is defined as the projection of total curvature $\kappa=|d\boldsymbol{T}/ds|$ on tangent space
\begin{equation}
	\kappa_{g} = \frac{d\boldsymbol{T}}{ds} \cdot \boldsymbol{e_{2}} 
	= \frac{d^{2}\boldsymbol{r}(s)}{ds^{2}} \cdot \boldsymbol{e_{2}} 
\end{equation}
Furthermore, the normal curvature $\kappa_{n}$ is defined as the projection of total curvature $\kappa=|d\boldsymbol{T}/ds|$ on normal vector $\boldsymbol{n}$
\begin{equation}
	\kappa_{n} = \frac{d\boldsymbol{T}}{ds} \cdot \boldsymbol{n} 
	= \frac{d^{2}\boldsymbol{r}(s)}{ds^{2}} \cdot \boldsymbol{e_{3}} 
\end{equation}
The geodesic curvature $\kappa_{g}$ depends only on the intrinsic properties of surface $S$, while the normal curvature $\kappa_{n}$ relies on the embedding of surface $S$ into three dimensional Euclidean space. Using geodesic curvature and normal curvature, the total curvature of a curve $\gamma(s)$ in three dimensional Euclidean space can be expressed by the square root $\kappa=\sqrt{\kappa_{g}^{2}+\kappa_{n}^{2}}$. In particular, the curves with zero geodesic curvature ($\kappa_{g}=0$) are viewed as ``straight line'' in surface $S$, and they are called geodesics in surface $S$. The curves with zero total curvature ($\kappa=\sqrt{\kappa_{g}^{2}+\kappa_{n}^{2}}=0$) are exactly the straight line (geodesics) in three dimensional Euclidean space.

\begin{table*}
	\caption{This table summarizes the features of geodesic curvature $\kappa_{g}$, total curvature $\kappa$ and Gauss curvature $\mathcal{K}$, as well as the differences between them.}
	\label{table3}
	\vspace{2mm}
	\begin{ruledtabular}
		\begin{tabular}{lllll}
			& Geodesic Curvature & Total Curvature & Gauss Curvature 
			\\
			\hline
			Notation & $\kappa_{g}$ & $\kappa=\sqrt{\kappa_{g}^{2}+\kappa_{n}^{2}}$ & $\mathcal{K}=\mathcal{K}_{1}\mathcal{K}_{2}$
			\\
			\hline
			Meaning & Curvature of a curve $\gamma$ (viewed  & Curvature of a curve $\gamma$ (embedded  & Intrinsic curvature of a two dimensional 
			\\
			& in two dimensional surface $S$) & into higher dimensional space \footnote{In this supplemental material, we choose the embedding of surface $S$ into three dimensional Euclidean space as an example to show the connection between geodesic curvature and total curvature.}) & surface $S$
			\\
			\hline
			Property & It measures how far this & It measures how far this & It measures how far this two dimensional 
			\\
			& curve is from being a geodesic \ & curve is from being a geodesic \ & surface $S$ is from being a flat space 
			\\
			& in two dimensional surface $S$. & in higher dimensional space. & intrinsically. 
			\\
			\hline
			Intrinsic/Extrinsic & $\kappa_{g}$ is intrinsic quantity of $S$ & $\kappa$ is extrinsic quantity of $S$ & $\mathcal{K}$ is intrinsic quantity of $S$ &
			\\
			& (It is independent of the & (It depends on the embedding & (It is independent of the embedding of 
			\\
			&  embedding of surface $S$ into & of surface $S$ into higher & surface $S$ into higher dimensional space.)
			\\
			& higher dimensional space.) & dimensional space.) &
			\\
			\hline
			Zero Curvature & $\kappa_{g}=0 \ \Leftrightarrow\ $ $\gamma$ is a geodesic & $\kappa=0 \ \Leftrightarrow\ $ $\gamma$ is a geodesic & $\mathcal{K}=0 \ \Leftrightarrow\ $ the two dimensional surface
			\\
			Condition & in two dimensional surface $S$ & in higher dimensional space & $S$ is intrinsically flat \footnote{However, the surface $S$ may exhibit nonzero “extrinsic curvature” when embedded into higher dimensional space.}
		\end{tabular}
	\end{ruledtabular}
\end{table*}

The normal curvature $\kappa_{n}$ is the extrinsic quantity of surface $S$, which depends on the embedding into higher dimensional space. If we change the direction of $\gamma(s)$ in three dimensional space, the normal curvature of this curve at point $p$ also changes (we assume the curve $\gamma=\gamma(s)$ always pass through point $p$ during the variation). In some circumstances, the normal curvature $\kappa_{n}$ could reach its maximum and minimum. These extreme values could reflect the geometric properties of two dimensional surface $S$, and they are named as the principle curvatures of this surface at point $p$.
\begin{equation}
	\mathcal{K}_{1} = \text{max}\{\kappa_{n}\} 
	\ \ \ \ \ 
	\mathcal{K}_{2} = \text{min}\{\kappa_{n}\}
\end{equation}
The Gauss curvature of surface $S$ (at point $p$) is defined to be the product of principal curvatures \cite{Carmo1976,ChernWH}
\begin{equation}
	\mathcal{K} \equiv \mathcal{K}_{1}\mathcal{K}_{2}
\end{equation}
An elegant theorem in surface theory (the Gauss Theorema Egregium) states that: although principal curvatures $\mathcal{K}_{1}$ and $\mathcal{K}_{2}$ depend on the embedding of $S$ in higher dimensional space, the Gauss curvature $\mathcal{K} = \mathcal{K}_{1}\mathcal{K}_{2}$ is an intrinsic geometric quantity of surface $S$ regardless of embedding \cite{Berger}. 
The Gauss curvature becomes zero if and only if the two dimensional surface is intrinsically flat. The features of geodesic curvature $\kappa_{g}$, total curvature $\kappa$ and Gauss curvature $\mathcal{K}$ are summarized in table \ref{table3}.

As we have emphasized in table \ref{table3}, both geodesic curvature and Gauss curvature are intrinsic geometric quantities of two dimensional surface $S$, they can be calculated from the intrinsic metric of surface $S$ naturally. If we assign the curving linear coordinates $(x_{1},x_{2})$ for two dimensional surface $S$ such that its intrinsic metric is expressed as
\begin{equation}
	ds^{2}=g_{11}dx_{1}^{2}+g_{22}dx_{2}^{2}
\end{equation}
for any curve $\gamma=\gamma(s)=(x_{1}(s),x_{2}(s))$ lived in this surface $S$ with arc length parameter $s$, its geodesic curvature can be calculated through the following equation
\begin{equation}
	\kappa_{g}(\gamma) = \frac{d\alpha}{ds} 
	- \frac{1}{2\sqrt{g_{22}}}\frac{\partial \ log(g_{11})}{\partial x_{2}} \cos\alpha
	+ \frac{1}{2\sqrt{g_{11}}}\frac{\partial \ log(g_{22})}{\partial x_{1}} \sin\alpha
\end{equation}
Here, $\alpha$ is the angle between tangent vector of $\gamma$ and coordinate axis $x_{1}$. Since geodesic curvature measures how far a curve is from being a geodesic in two dimensional surface $S$, the vanishing of geodesic curvature indicates that $\gamma=\gamma(s)$ is a geodesic curve in this surface.
\begin{equation}
	\kappa_{g}(\gamma)=0 
	\ \Leftrightarrow \ 
	\nabla_{\boldsymbol{T}}\boldsymbol{T} \bigg|_{\gamma=\gamma(s)} = 0
	\ \Leftrightarrow \ 
	\bigg[ \frac{d^{2}x_{i}}{ds^{2}}-\Gamma_{jk}^{i}\frac{dx_{j}}{ds}\frac{dx_{k}}{ds} \bigg]_{\gamma=\gamma(s)} = 0
\end{equation}
The Gauss curvature of two dimensional surface $S$ is calculated by
\begin{equation}
	\mathcal{K} = \mathcal{K}_{1}\mathcal{K}_{2}
	= \frac{R_{1212}}{g_{11}g_{22}-(g_{12}^{2})}
	= -\frac{1}{\sqrt{g_{11}g_{22}}}
	\bigg[
	\frac{\partial}{\partial x_{2}} \bigg( \frac{1}{\sqrt{g_{22}}} \frac{\partial\sqrt{g_{11}}}{\partial x_{2}}  \bigg)
	+ \frac{\partial}{\partial x_{1}} \bigg( \frac{1}{\sqrt{g_{11}}} \frac{\partial\sqrt{g_{22}}}{\partial x_{1}}  \bigg)
	\bigg] \label{Gauss-Curvature1}
\end{equation}
where $R_{1212}$ is the Riemannian curvature tensor of this surface. In this supplemental material, we adopt the notation in surface theory, where covariant and contravariant coordinates are not distinguished. 

In the present work, we focus on the optical geometry of spherically symmetric black holes. If we restrict test particles moving in equatorial plane $\theta=\pi/2$ and assign the coordinates $(x_{1},x_{2})=(r,\phi)$, then the two dimensional optical metric becomes 
\begin{equation}
	dt^{2}=\tilde{g}^{\text{OP}}_{ij}dx^{i}dx^{j}
	=\frac{1}{[f(r)]^{2}}dr^{2}+\frac{r^2}{f(r)}d\phi^{2}
	\label{optical metric1}
\end{equation}
Here $t$ is the arc length parameter / spatial distance parameter in optical geometry. For any circular photon orbit, the angle between this orbit and coordinate axis $r$ is $\alpha=\pi/2$, then the geodesic curvature $\kappa_{g}$ of circular photon orbit $r=r_{ph}$ reduces to
\begin{equation}
	\kappa_{g}(r=r_{ph}) =  
	\bigg[ \frac{d\alpha}{dt} 
	- \frac{1}{2\sqrt{\tilde{g}^{\text{OP}}_{\phi\phi}}} \frac{\partial \ log(\tilde{g}^{\text{OP}}_{rr})}{\partial \phi} \cos\alpha
	+ \frac{1}{2\sqrt{\tilde{g}^{\text{OP}}_{rr}}} \frac{\partial \ log(\tilde{g}^{\text{OP}}_{\phi\phi})}{\partial r} \sin\alpha \bigg]_{r=r_{ph}} 
	=  \bigg[
	\frac{1}{2\sqrt{\tilde{g}^{\text{OP}}_{rr}}} \frac{\partial \ log(\tilde{g}^{\text{OP}}_{\phi\phi})}{\partial r}
	\bigg]_{r=r_{ph}}   
\end{equation}
Using this coordinates $(x_{1},x_{2})=(r,\phi)$, the Gauss curvature of two dimensional optical geometry in equation (\ref{Gauss-Curvature1}) becomes
\begin{equation}
	\mathcal{K}(r,\phi) =  -\frac{1}{\sqrt{\tilde{g}^{\text{OP}}}}
	\bigg[
	\frac{\partial}{\partial \phi} \bigg( \frac{1}{\sqrt{\tilde{g}^{\text{OP}}_{\phi\phi}}} \frac{\partial\sqrt{\tilde{g}^{\text{OP}}_{rr}}}{\partial \phi}  \bigg)
	+ \frac{\partial}{\partial r} \bigg( \frac{1}{\sqrt{\tilde{g}^{\text{OP}}_{rr}}} \frac{\partial\sqrt{\tilde{g}^{\text{OP}}_{\phi\phi}}}{\partial r}  \bigg)
	\bigg] 
\end{equation}
where $\tilde{g}^{\text{OP}}=det(\tilde{g}^{\text{OP}}_{ij})=\tilde{g}^{\text{OP}}_{rr}\tilde{g}^{\text{OP}}_{\phi\phi}$ is the determinant of optical metric. These results coincide with equation (5) and equation (6) in the main text exactly.

Furthermore, in spacetime geometry (which is a four dimensional Lorentz manifold), one can pick two dimensional surface $S_{\mu\nu}$ with two tangent vector fields $\partial_{\mu}$ and $\partial_{\nu}$. For each of these two dimensional surfaces, the Gauss curvature and geodesic curvature can be defined in similar ways. Particularly, the definition of Gauss curvature is similar to the expression in equation (\ref{Gauss-Curvature1}), namely via $\mathcal{K}=\frac{R_{\mu\nu\mu\nu}}{g_{\mu\mu}g_{\nu\nu}-(g_{\mu\nu})^{2}}$ \cite{Jost2011}. In this way, a collection of Gauss curvatures can be obtained in a four dimensional Lorentz manifold. Mathematicians often call these Gauss curvatures (for different two dimensional surfaces $S_{\mu\nu}$) as sectional curvatures of Lorentz manifold \cite{Jost2011,Lee2019}. 



